# Anharmonicity in multifrequency atomic force microscopy

Sergio Santos and Victor Barcons

Departament de Disseny i Programació de Sistemes Electrònics, UPC - Universitat Politècnica de Catalunya, Av. Bases, 61, 08242 Manresa (Barcelona), Spain

## Abstract

In multifrequency atomic force microscopy higher eigenmodes are externally excited to enhance resolution and contrast while simultaneously increasing the number of experimental observables with the use of gentle forces. Here, the implications of externally exciting multiple frequencies are discussed in terms of cantilever anharmonicity, fundamental period and the onset of subharmonic and superharmonic components. Cantilever anharmonicity is shown to affect and control both the observables, that is, the monitored amplitudes and phases, and the main expressions quantified via these observables, that is, the virial and energy transfer expressions which form the basis of the theory.

## I. INTRODUCTION

Over the past decades the field of atomic force microscopy AFM has evolved from quasi-static[5] to dynamic methods[3, 18, 39] and more recently to dynamic methods where multiple frequencies are externally excited. In standard monomodal AFM the cantilever is externally excited at a single frequency[28]. Then, when the tip interacts with the nonlinear tip-sample forces, exact multiples of the drive frequency, i.e. its higher harmonics, are excited[28, 33]. The amplitude, phase and mean deflection of the response at the frequency of the drive are the experimental observables. Other than in heavily damped environments[4], and provided the tip-sample forces are gentle[24], higher harmonics typically lie too close to the noise level and are neglected[8, 26, 35]. In the multifrequency approach two[19] or more[15, 32] external drives are employed to excite the cantilever at or near the natural frequencies or eigenmodes of the free cantilever. This gives rise to the simultaneous detection of multiple experimental observables and secondary contrast channels[10] that might lead to enhanced resolution[14] while providing the means to robustly quantify sample properties[13]. Still, and while attractive for the development of the field, the simultaneous excitation of multiples frequencies at or near the eigenmodes comes at the cost of additional instrumentation [22], added complexity to cantilever dynamics[2, 34] and the requirement of interpreting secondary contrast channels[6, 15, 29]. Commercial cantilevers are generally anharmonic in the sense that the higher natural frequencies are not multiples of the fundamental[22, 27]. Such anharmonicity can be derived analytically from standard beam theory[36]. While the presence of the tip might lead to slight deviations[9] in real applications, in general, the commonly employed commercial cantilevers remain anharmonic[22, 31]. Here, the consequences of cantilever anharmonicity/harmonicity are discussed in terms of external drive frequencies, fundamental period of oscillation, eigenmodes and the emerging general analytical theory in the field of multifrequency AFM.

## II. THEORY AND SIMULATIONS

The general response of a driven cantilever that follows periodic motion with period T can be written as[7]

$$z = z_0 + \sum_{n>0} z_n = z_0 + \sum_{n>0} A_n \sin(n\omega_{SH} t + \phi_n) \quad (1)$$

where z stands for tip position in the direction normal to the surface of the unperturbed cantilever, $z_0$ is the mean deflection and $z_n$ is the $n_{th}$ harmonic component of motion. The fundamental period T can be associated with a fundamental frequency $\omega_{SH}=2\pi f_{SH}$ where $\omega_{SH}$ is the fundamental angular frequency, $f_{SH}$ is the corresponding frequency in Hz and SH stands for subharmonics in this work throughout; $\omega_{SH}$ and $f_{SH}$ are used interchangeably to refer to fundamental frequency. In multifrequency AFM several external drives are employed to excite the cantilever at frequencies near the modal resonant frequencies $\omega_{(m)}$; m stands for mode number and mode numbers are bracketed in this work to distinguish them from harmonic number n as done elsewhere[29, 30]. For simplicity and without loss of generality we reduce the cantilever to two modes and employ two external drives as in standard bimodal AFM[14, 23, 32]. The drive ($\omega_{D1}$ and $\omega_{D2}$) and cantilever ($\omega_{(1)}$ and $\omega_{(1)}$) frequencies are used interchangeably from now on without causing ambiguity since $\omega_{D1} \approx \omega_{(1)}$ and $\omega_{D2} \approx \omega_{(2)}$; D1 and D2 stand for drive 1 and 2 respectively. Furthermore, higher harmonics of $\omega_{SH}$ are here termed superharmonics or simply subharmonics from now on. These refer to frequencies that are a fraction of $\omega_{(1)}$ or, in general, multiples of $\omega_{SH}$ but not necessarily multiples of the external drive frequencies. Then

$$\omega_{SH} = \frac{\omega_{D1}}{p} = \frac{\omega_{D2}}{q} \qquad \text{where r=q/p} \tag{2}$$

$$T = \frac{2\pi}{\omega_{SH}} \tag{3}$$

where p and q are integer numbers that, according to the Fourier theorem[38], have no common multiples, i.e. r=q/p cannot be further reduced. An implication of the above discussion is that the user's choice of $\omega_{D1}$ and $\omega_{D1}$ (or choice of cantilever with a given relationship between $\omega_{(1)}$ and $\omega_{(2)}$) dictates the numerical value of the fundamental frequency $\omega_{SH}$ and hence the fundamental period T. Similar phenomena was already noticed by Basak and Raman when discussing experimental and numerical results in monomodal AFM operated in liquid environments[4], but higher harmonics of the fundamental drive only were discussed. Here we point out that such phenomena is general in multifrequency AFM. That is, physically, analytical derivations relying on steady-state oscillation and the principle of conservation of energy are valid provided the frequency $\omega_{SH}$, and not $\omega_{D1}$ or $\omega_{D2}$, is considered as fundamental. Note that the theory of multifrequency AFM[16, 17] is based on the concepts of virial[21] and energy dissipation[8] that rely on the above principles. We now define cantilever anharmonicity/harmonicity according to whether there exists an integer r=q/p such that $\omega_{(2)}=r\omega_{(1)}$. In terms of p and q in (2) harmonicity implies that p=1 and q=r ($\omega_{SH}=\omega_{D1}$ and $\omega_{D2}=q\omega_{D1}$). Anharmonicity implies that p>1 and q≠r (p$\omega_{SH}=\omega_{D1}$ and $\omega_{D2}=q\omega_{SH}$). In terms of the fundamental period a modal virial $V_{(m)}$ can be defined [29] as

$$V_{(m)} = \frac{1}{T} \int_{t=t'}^{t=t'+T} F_{ts} z_{(m)} dt \tag{4}$$

where T is given by (3), $F_{ts}$ is the unrestricted tip-sample force and $z_{(m)}$ is the modal tip position. Next we reduce the cantilever to the two first fundamental modes (M=2 where m=1 and 2). The equations of motion are

$$\frac{k_{(m)}}{\omega_{(m)}^2}\ddot{z}_{(m)}(t) + \frac{k_{(m)}}{Q_{(m)}\omega_{(m)}}\dot{z}_{(m)}(t) + k_{(m)}z_{(m)} = F_{0D1}\sin(\omega_{D1}t) + F_{0D2}\sin(\omega_{D2}t) + F_{ts} \qquad (5)$$

where $k_{(m)}$, $Q_{(m)}$, $\omega_{(m)}$ and $z_{(m)}$ are the spring constant, quality factor and natural frequency of the m eigenmode. $F_{0D1} \equiv F_{0(1)}$ and $F_{0D2} \equiv F_{0(2)}$ are the magnitudes of the two external drive forces acting at or near $\omega_{(1)}$ and $\omega_{(2)}$ respectively. The modal position $z_{(m)}$ can be expanded in terms of higher harmonics

$$z_{(m)} \approx z_{(m)0} + \sum_{n>0}^{N} A_{(m)n}\sin(n\omega_{SH}t + \phi_{(m)n}) \qquad (6)$$

where $z_{(m)0}$ is the mean deflection of $z_{(m)}$ and the harmonic amplitudes and phases of mode m and harmonic n are written as $A_{(m)n}$ and $\phi_{(m)n}$ respectively. Combining (4) to (6) leads to

$$V_{(m)} = -\frac{1}{2}F_{0(m)}A_{(m)n}\cos\phi_{(m)n} + HD(V_{(m)}) \qquad (7)$$

where the harmonic number n for the first term on the right is to be replaced by p or q when m=1 and m=2 respectively. $HD(V_{(m)})$ is defined as the harmonic distortion of the modal virial $V_{(m)}$. Furthermore, a modal term $E_{T(m)}$[20] that has been recently associated[29] with a combination of energy transfer between modes and irreversible loss of energy in the tip sample interaction can be defined as

$$E_{T(m)} = -\int_{t=t'}^{t=t'+T} F_{ts}\dot{z}_{(m)}\,dt \qquad (8)$$

$E_{T(m)}$ is here termed modal Energy Transfer. Combining (5) and (6) with (8) leads to

$$E_{T(m)} = \pi n F_{0(m)} A_{(m)n} \sin\phi_{(m)n} + \pi \frac{k_{(m)}}{Q_{(m)}} \frac{\omega_{SH}}{\omega_{(m)}} n^2 A_{(m)n}^2 + HD(E_{T(m)}) \qquad (9)$$

where the harmonic number n is again to be replaced by p or q for m=1 and m=2 as in (7). $HD(E_{T(m)})$ is also associated with harmonic distortion but in this case of modal energy transfer $E_{T(m)}$. The irreversible loss of energy $E_{dis}$ in the tip-sample interaction per fundamental period T is

$$E_{dis} = \sum_{m>0} E_{T(m)} \qquad (10)$$

Finally, the Virial $V_{(m)n}$ and energy transfer $E_{T(m)n}$ associated with the monitored frequencies n=p and q (and modes m=1 and m=2 respectively) are

$$V_{(m)n} \approx V_n = -\frac{1}{2} F_{0n} A_n \cos\phi_n \qquad (11)$$

$$E_{T(m)n} \approx E_{Tn} = \pi n \frac{k_{(m)} A_{0n} A_n}{Q_{(m)}} \left[ \sin\phi_n - \frac{A_n}{A_{0n}} \frac{n\omega_{SH}}{\omega_{(m)}} \right] \qquad (12)$$

where $A_{0n}$ is the free amplitude for the unperturbed cantilever at the given harmonic n=p or q. The monitored phase shifts $\phi_p$ (n=p) and $\phi_q$ (n=q) for m=1 and m=2 follow from (11) and (12)

$$\phi_n \approx \tan^{-1}\left[ -\frac{\frac{E_{Tn} Q_{(m)}}{2\pi n} + \frac{1}{2} k_{(m)} A_n^2}{V_n Q_{(m)}} \right] \qquad (13)$$

The expression in (13) can be computed in terms of experimental observables and is equivalent to others proposed in the literature[16, 29]. Nevertheless an important remark is made here in terms of anharmonicity/harmonicity and fundamental period T. Namely, the expressions for $V_{(m)n}$ (11) and $E_{T(m)n}$ (12) are to be computed over a fundamental (subharmonics) period T as given by (3). Furthermore, the amplitudes $A_n$ and phases $\phi_n$ are to be averaged over the fundamental period T. The validity of these statements is next discussed with the use of results obtained via numerical integration of (5).

In the simulations and in the long range $F_{ts}$ is defined by the Hamaker constant H, the effective tip radius R and the tip sample distance d[21]

$$F_{ts}(d) = -\frac{RH}{6d^2} \qquad\qquad a_0 < d \qquad\qquad (14)$$

where $a_0$ (≈0.165 nm) is an intermolecular distance. In the short range $d \leq a_0$ $F_{ts}$ is modeled by[11, 21]

$$F_{ts}(d) = -\frac{RH}{6a_0^2} + \frac{4}{3}E^*\sqrt{R}\delta^{3/2} \qquad\qquad a_0 \geq d \qquad\qquad (15)$$

where d and the tip-sample deformation $\delta$ are related by $\delta = a_0 - d$. $E^*$ (≈1 GPa in this work) is the effective Young modulus in the contact. Numerical results for anharmonic and harmonic cantilevers are presented in Fig. 1. The generic parameters are: $A_{0p}$= 20 nm, $A_p$= 10 nm, $k_{(1)}$=2 N/m, $k_{(2)}$=80 N/m, $Q_{(1)}$=100, $Q_{(2)}$=600, H= 2.1×10$^{-20}$ J and R=5 nm. These conditions have led to the average force remaining positive throughout, i.e. the cantilever oscillated in the repulsive regime[11].

First, an anharmonic cantilever has been selected for which $f_{(1)}$ = 70 kHz ($\omega_{(1)}$=2π$f_{(1)}$) and $f_{(2)}$= 443.1 kHz ($\omega_{(2)}$=2π$f_{(2)}$). Such anharmonic cantilevers are standard from commercial suppliers[22, 27, 31]. The fundamental (subharmonics) frequency follows from (2) and (3) giving $\omega_{(2)}/\omega_{(1)}$=q/p=633/100=r=6.33. The subharmonics frequency is $f_{SH}$=0.7 kHz. The physical implication is that the frequency spectrum of such a cantilever when interacting with the surface should produce peaks separated by integer multiples of 0.7 kHz. This hypothesis is confirmed by looking at the results in Fig. 1a (and zoom in Fig. 1b). The larger subharmonics/superharmonics in Figs. 1a and 1b lie in the order ~10 pm while the smaller ones can be as small as fractions of a fm. It is also interesting to note that by lacking resolution Figs. 1a and 1b might lead to believe that the fundamental (subharmonics) frequency was 70/3 kHz. Importantly, the frequency corresponding to n=q, i.e. the drive frequency near the second mode, is a higher harmonic of the fundamental subharmonics frequency and not of the drive frequency of the first mode. Still, other than for n=q, the numerical values show that multiples drive frequency near the second mode, i.e. the standard higher harmonics, are larger in magnitude than the superharmonics or multiples of the fundamental subharmonics. Second, an anharmonic cantilever has been selected for which $f_{(1)}$ = 70 kHz ($\omega_{(1)}=2\pi f_{(1)}$) and $f_{(2)}$= 434 kHz ($\omega_{(2)}=2\pi f_{(2)}$). In this case $\omega_{(2)}/\omega_{(1)}$=q/p=31/5=r=6.2 . The corresponding subharmonics frequency is $f_{SH}$=14 kHz. The results are shown in Figs. 1c and 1d where 1d is a zoom of 1c. Again subharmonics amplitudes can reach values of ~10pm. Finally a harmonic cantilever is selected for which $f_{(1)}$ = 70 kHz ($\omega_{(1)}=2\pi f_{(1)}$) and $f_{(2)}$= 420 kHz ($\omega_{(2)}=2\pi f_{(2)}$). In this case $\omega_{(2)}/\omega_{(1)}$=q/p=6/1=r=6 produces a subharmonics frequency of $f_{SH}$=70 kHz. That is, the fundamental of subharmonics frequency coincides with $\omega_{(1)}$. In this case the results in Fig. 1e (and zoom in 1f) show that only harmonics of $\omega_{(1)}$ are excited. This is the standard assumption. The conclusions in terms of the fundamental period T are that for r=6.33 (Figs. 1a and 1b), r=6.2 (Figs. 1c and 1d) and r=6 (Figs. 1e and 1f), T≈1,400 μs, T≈71 μs and T≈14 μs respectively. This imposes limitations in terms of scanning rate depending on r or anharmonicity in multifrequency AFM as discussed below.

We note that when the fundamental theory of multifrequency AFM was developed[16], subharmonics peaks were already observed but were assumed to be too small to cause significant perturbation. Nevertheless the tendency of employing small oscillation amplitudes for secondary contrast channels implies that there are cases when these peaks should not be underestimated. More thoroughly, the two basic expressions in dynamic AFM, i.e. the virial (4) and the energy dissipation or energy transfer (8), and the main observables, i.e. amplitudes and phases at n=p and n=q, can vary significantly during a fundamental subharmonics period. Moreover the presence of subharmonics might lead to transitions in operation regime according to the standard phase shift convention[11]. In order to quantify variations in observables some of the numerical results from the simulations in Fig. 1 are given in Table I. From those it is particularly interesting to note the variations in the sign of the virial $V_q$ in (11) according to the anharmonicity/harmonicity of the cantilever as parameterized by r. Since whether $\phi_q$ lies above or below 90 degrees depends on $V_q$, the choice of r affects where $\phi_q$ lies. Furthermore, the sign of the energy transfer from and to other frequencies to and from the monitored frequency q, i.e. $E_{Tq}$, also depends on r. All values in Table I, except the phase shift $\phi_q$ which is given in degrees, are given in eV. Finally note that $E_{dis}$=0.00 throughout since the interaction is conservative.

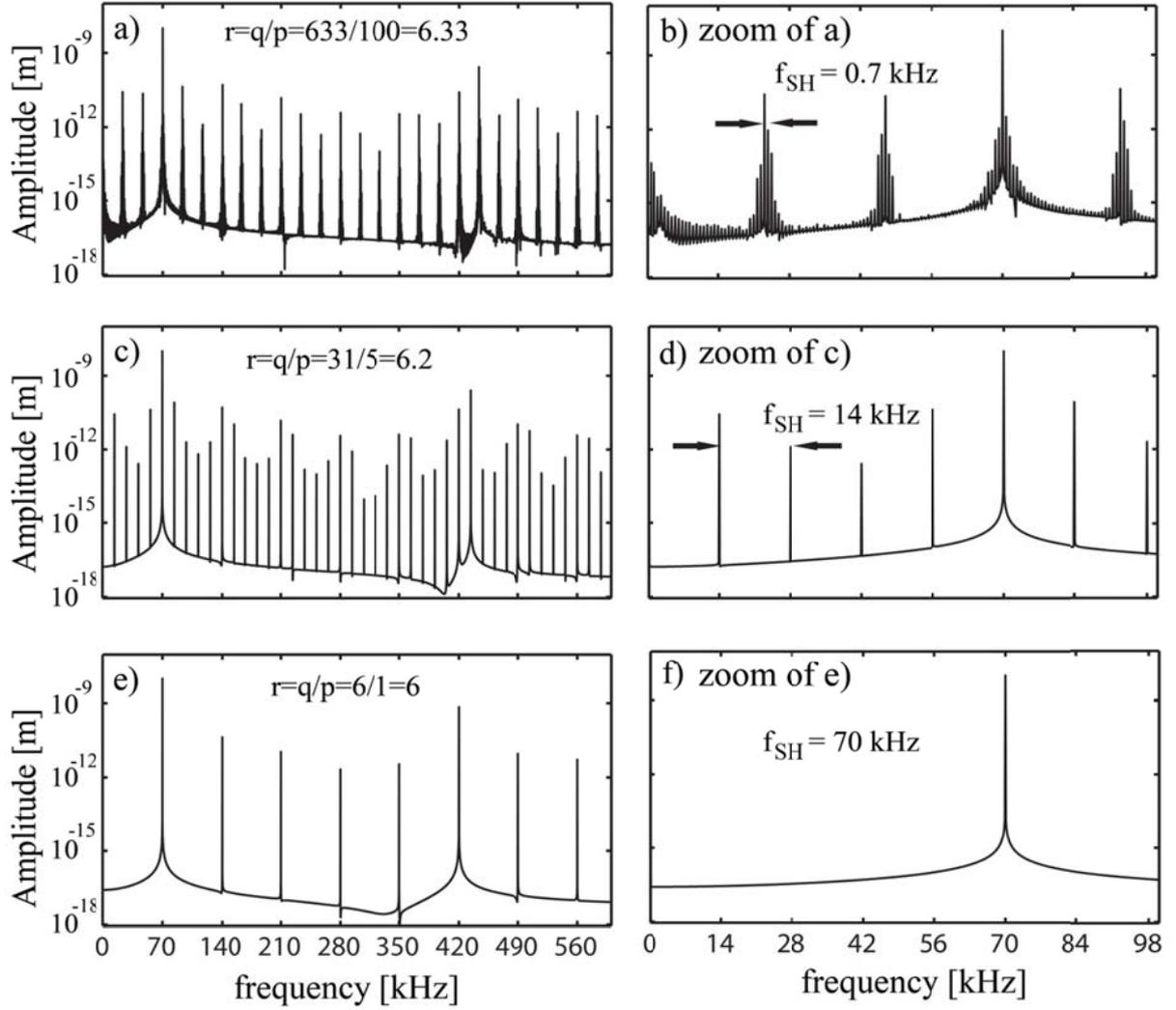

FIG. 1 Examples of the frequency spectrum (amplitude/frequency) for an interacting tip according to cantilever anharmonicity/harmonicity as parameterized by r=q/p. The panels on the right correspond to zooms of the panels on the left.

|  | Eq. (4) | Eq. (4) | Eq. (8) | Eq. (8) | Eq. (10) | Eq. (11) | Eq. (12) | Eq. (13) | (Sim) |
|---|---|---|---|---|---|---|---|---|---|
|  | $V_{(1)}$ [eV] | $V_{(2)}$ [eV] | $E_{T(1)}$ [eV] | $E_{T(2)}$ [eV] | $E_{dis}$ [eV] | $V_q$ [eV] | $E_{Tq}$ [eV] | $\Phi_q$ [°] | $\Phi_q$ [°] |
| r=633/100 | -10.82 | -0.07 | 3.55 | -3.55 | 0.00 | -0.11 | -0.29 | 15.88 | 15.88 |
| r=31/5 | -10.83 | -0.05 | 0.23 | -0.23 | 0.00 | -0.10 | 0.03 | 14.80 | 14.81 |
| r=6/1 | -9.53 | -0.31 | 11.29 | -11.29 | 0.00 | 0.32 | -7.61 | 171.32 | 171.32 |

Table I. Numerical values of some of the expressions derived in this work as obtained in the simulations discussed in Fig. 1. The equation numbers are specified and the numerical values were obtained from an FFT. The last column corresponds to the value of phase shift for n=q (monitored phase shift near the second mode) resulting from simulations (Sim).

At this point we note that the frequency spectrum plots in Fig. 1 have been obtained by employing the standard FFT algorithm with sufficient resolution to detect the fundamental subharmonics frequency and its higher harmonics or superharmonics. The values in Table I were computed from the numerical values obtained directly from the FFT. Nevertheless, in standard AFM equipment lock-in-amplifiers are typically employed. In particular, standard expressions in AFM are commonly derived by manipulating the equation of motion (5) and integrating over a period $T_{(1)}$; $T_{(1)}$ stands for a period relative to the external drive close to mode one, that is a cycle of the first external drive. In this way

$$T_{(1)} = \frac{2\pi}{\omega_{(1)}} \tag{16}$$

One could similarly define a period $T_{(2)}$ for mode 2. The tip position is then approximated[22, 23, 34] to

$$z \approx A_p \sin(\omega_{D1} t + \phi_p) + A_q \sin(\omega_{D2} t + \phi_q) \tag{17}$$

where harmonics n=p and q are accounted for as the main contributions from modes 1 and 2 respectively. Note that $A_p$, $A_q$, $\phi_p$ and $\phi_q$ are typically termed $A_1$, $A_2$, $\phi_1$ and $\phi_2$ respectively. Lock-in-amplifiers however tend to average amplitudes and phases over c cycles of the monitored frequency. This implies an averaging of the expressions that can be computed with these observables such as viral $V_{(m)}$ and energy transfer $E_{T(m)}$. If averaging over c cycles the effective modal virial and energy transfer can be computed as

$$<V_{(m)}> \equiv \frac{1}{c} \sum_{x=1}^{x=c} \frac{1}{T_{(1)}} \int_{t=t'}^{t=t'+T_{(1)}} F_{ts} z_{(m)} dt(x) \tag{18}$$

and

$$<E_{T(m)}> \equiv \frac{1}{c}\sum_{x=1}^{x=c} - \int_{t=t'}^{t=t'+T_{(1)}} F_{ts}\, \dot{z}_{(m)}\, dt(x) \quad (19)$$

where x is simply an index that runs over c cycles. Regarding modal and single frequency expressions, it is already clear from the numerical values in Table I that there are significant differences in value and even sign. Compare for example to modal virial (4) for m=2 with the single frequency virial in (11) for q in Table I. This is in agreement with very recent findings[29] and implies that modal and single frequency virial at the monitored frequencies should not be equated.  Note that while the next discussion focuses on modal values it is also relevant to the single frequency (monitored) virial and energy transfer (not shown). Several interesting outcomes follow. First note that the main difference between the definitions of the averages in (18) and (19) and the actual modal viral $V_{(m)}$ and energy transfer $E_{T(m)}$ in (4) and (8) is that the integrals are carried out over different periods of oscillation, i.e. T from (3) and $T_{(m)}$ from (16) respectively.  This might lead to errors depending on the anharmonicity of the cantilever, as parameterized by r in (2), and the number of cycles c over which the signals are averaged in (18) and (19). Errors in $V_{(m)}$ and $E_{T(m)}$ are shown in Table II for the same values of r as in Fig. 1 and Table I and by employing the same cantilever-operational parameters. In particular, the values r=6.33, r=6.2 and r=6 corresponding to p=100, 5 and 1 have been employed. The results of averaging over c=1, 5 and 100 are shown in the table for $V_{(m)}$ and $E_{T(M)}$. The integration has been carried out by employing the period $T_{(1)}$ in (16) throughout. Errors in the table are shown as percentages. The main outcome is that the error is zero throughout only when the cantilever is harmonic, i.e. r is an integer, as predicted. Errors for unharmonic cantilevers are obtained only when averaging over c=p or multiples of p. This is in agreement with the discussion above since averaging over c=p is equivalent to employing the fundamental (subharmonics) frequency in (3). The values for the harmonic distortion in virial and energy transfer are also given in the table for the different values of anharmonicity r and averaging periods c. These errors are given as percentages in Table II and are computed as

$$\overline{HD(V_{(m)})} \equiv \frac{V_{(m)} - V_n}{V_{(m)}} \qquad (20)$$

$$\overline{HD(E_{T(m)})} = \frac{E_{T(m)} - E_{Tn}}{E_{T(m)}} \qquad (21)$$

From the results in Table II it is important to emphasize that conservative interactions are typically computed from the virial expression[12, 25]. Dissipative interactions on the other hand are accounted for by the energy transfer expression[8, 37]. More complex expressions and combinations result for the conservative and dissipative interactions when higher harmonics are accounted for[1, 9]. Nevertheless, in general, errors in the numerical computation of the virial and energy transfer will lead to errors in quantification. Thus, anharmonicity as parameterized by r in (3) and possible averaging over periods other than the fundamental, i.e. expressions such as (18) and (19), should be considered with care. Overall, when simultaneously exciting multiple frequencies externally, amplitudes and phases resulting from lock-in-amplifiers should be compared with those resulting from an FFT obtained with sufficient resolution as to resolve possible subharmonics frequencies. Finally, it should be noted that anharmonicity could be controlled by designing cantilevers with the appropriate geometry as already discussed by Sahin et al.[27]. Otherwise, the relationship between drive frequencies could be carefully selected by the user when employing standard anharmonic cantilevers in order to reduce anharmonicity and the onset of subharmonics. This simply implies selecting frequencies that reduce p and q in (2) while driving near the natural cantilever frequencies. We also note that the discussion in this work also affects the cantilever operated in the attractive regime (not shown).

|  | Eq. (18) c=1 | Eq. (18) c=5 | Eq. (18) c=100 | Eq. (19) c=1 | Eq. (19) c=5 | Eq. (19) c=100 | Eq. (20) HD($V_{(1)}$) | Eq. (20) HD($V_{(2)}$) | Eq. (21) HD($E_{T(1)}$) | Eq. (21) HD($E_{T(2)}$) |
|---|---|---|---|---|---|---|---|---|---|---|
| r=633/100 | 23.29 | 4.57 | 0.00 | 23364.42 | 4666.09 | 0.00 | 0.12 | -61.01 | 115.70 | 91.75 |
| r=31/5 | 26.58 | 0.00 | 0.00 | 15943.16 | 0.00 | 0.00 | 0.18 | -124.96 | 83.31 | 114.94 |
| r=6/1 | 0.00 | 0.00 | 0.00 | 0.00 | 0.00 | 0.00 | -0.24 | 201.67 | 0.50 | 32.64 |

Table II. Errors in virial and energy transfer resulting from averaging over c cycles for a range of cantilever anharmonicity parameterized by r. The errors are given as percentages.

### III. CONCLUSIONS

In summary, the implications of employing anharmonic cantilevers in multifrequency AFM have been discussed regarding periodicity, fundamental period and the onset of subharmonic and superharmonic components. Anharmonicity affects both the observables, that is, the monitored amplitudes and phases, and the main expressions resulting from observables, that is, the virial and energy transfer expressions which form the basis of the analytic theory. In general, the theory and results in this work should lead to improvements in the reduction of noise, interpretation of contrast and robustness in terms of quantification of properties in multifrequency atomic force microscopy.